# Experimental evidence for topological surface states wrapping around bulk SnTe crystal

K. Dybko,* M. Szot, A. Szczerbakow, M. U. Gutowska, T. Zajarniuk, J. Z. Domagala, A. Szewczyk, T. Story, and W. Zawadzki
*Institute of Physics, Polish Academy of Sciences, Aleja Lotnikow 32/46 PL-02668 Warsaw, Poland*

We demonstrate that the metallic topological surface states *wrap on all sides* the 3D topological crystalline insulator SnTe. This is achieved by studying oscillatory quantum magneto-transport and magnetization at tilted magnetic fields which enables us to observe simultaneous contributions from neighbouring sample sides. Taking into account pinning of the Fermi energy by the SnTe reservoir we successfully describe theoretically the de Haas-van Alphen oscillations of magnetization. The determined π-Berry phase of surface states confirms their Dirac fermion character. We independently observe oscillatory contributions of magneto-transport and magnetization originating from the bulk SnTe reservoir of high hole density. It is concluded that the bulk and surface Landau states exist in parallel. Our main result that the bulk reservoir is surrounded on all sides by the topological surface states has an universal character.

Topological insulators (TI) and topological crystalline insulators (TCI) are new phases of quantum matter with topologically protected gapless boundary states. The topological protection is ensured by time reversal symmetry (TI) or specific crystalline symmetry (TCI), respectively. It was predicted theoretically for a nontrivial band ordering of a bulk semiconductor that the two-dimensional (2D) topologically protected surface states appear on *all* surfaces of the bulk material [1, 2]. For two opposite surfaces this property was demonstrated experimentally for thin films of strained HgTe [3] and unstrained $BiSbTeSe_2$ [4]. A separation of the Shubnikov-de Haas oscillations originating from the two surfaces was achieved because they had different 2D electron densities [3]. The IV-VI compound SnTe, having the rock-salt crystal symmetry and the nontrivial band ordering at the L points of the Brillouin zone, is known to generate the topological states of the TCI at the (001), (110) and (111) surfaces. This has been first demonstrated in the angle resolved photoemission spectroscopy (ARPES) studies [5] and by scanning tunneling spectroscopy [6, 7].

In the present work we show that the topological surface states (TSS) appear equally well on neighbouring surfaces thus wrapping bulk SnTe sample. As compared to the strained HgTe, our system has two distinct novel properties. First, we deal with topological crystalline insulators whose topological protection is assured by the proper order and symmetry of energy bands in bulk SnTe, see [5, 8–10]. Second, we deal with a large reservoir of holes in the bulk SnTe which pins the Fermi level of the entire system and determines properties of TSS. We study the quantized Hall regime, the Shubnikov-de Haas effect and the de Haas-van Alphen effect. In order to separate contributions of the two investigated surfaces we rotate our sample with respect to the direction of magnetic field. In the studies of TI one usually tries to suppress the effect of the bulk by reducing its volume, carrier density etc. Very high quality of our samples allows us to separate contributions of the bulk and surface states without recoursing to any additional measures, so we deal with natural as-grown SnTe crystals [11].

The SnTe samples were cleaved along (100) and equivalent surfaces. The high crystal quality of our sample is illustrated in Fig. S1 of Supplemental Material [12]. The Hall-bar geometry with six contacts was used in transport experiments. Figure 1a shows raw data of magneto-resistivity tensor components measured for $B \parallel [001]$ probing the top and bottom surfaces. The results of $R_{xx}$ and $R_{xy}$, while showing slight wavy behavior due to SdH oscillations and contribution of the Quantum Hall Effect, do not exhibit the usual plateaus of $R_{xy}$ and zeros of $R_{xx}$. The Hall resistivity is several orders of magnitude lower than the expected kΩ range. This becomes understandable when one takes into account that the transport has two components: the surfaces and the bulk. The bulk of SnTe, having very large hole density, plays the role of a reservoir for the system. This reservoir short-circuits the surface conductivity. The role of reservoir in the magneto-conductivity of 2D semiconductor systems is known, see refs [13–15]. In particular, it was observed in agreement with our data that the reservoir dramatically lowers measured values of the Hall voltage in the quantum Hall regime [16, 17]. Another important effect of a large reservoir is to pin the Fermi level of the system.

To characterize the quantum behavior of our system we subtract smooth backgrounds of the data shown in Fig. 1a. The resulting oscillatory behavior, plotted in 1/B scale, is shown in the inset. The observed frequency of both $R_{xy}$ and $R_{xx}$ oscillations is 26 T and they are shifted with respect to each other by a constant phase factor 1/4 of the full period 2π. This frequency corresponds to $N_s = 6.2\times10^{11}$ $cm^{-2}$ of holes, according to the Lifshitz-Kosevich theory of SdH oscillations of 2D gas [18]. The observed phase factor satisfies the resistivity rule $R_{xx} \sim B \cdot dR_{xy}/dB$ commonly obeyed in the integer quantum Hall regime [19], indicating that we deal with 2D states, see Fig. S2 [12] . It can be seen that in the high-field region of $1/B < 0.11$ T$^{-1}$ there appear small-amplitude oscillations better seen in $B$ scale (see Fig. S3 [12]). Their frequency is roughly 322 T indicating that the oscillations originate from ellipsoids at the L points of 3D SnTe. This conclusion is additionally confirmed by angular dependence of the observed frequencies [12]. The above analysis clearly shows that our transport results represent a sum of surface and bulk contributions.

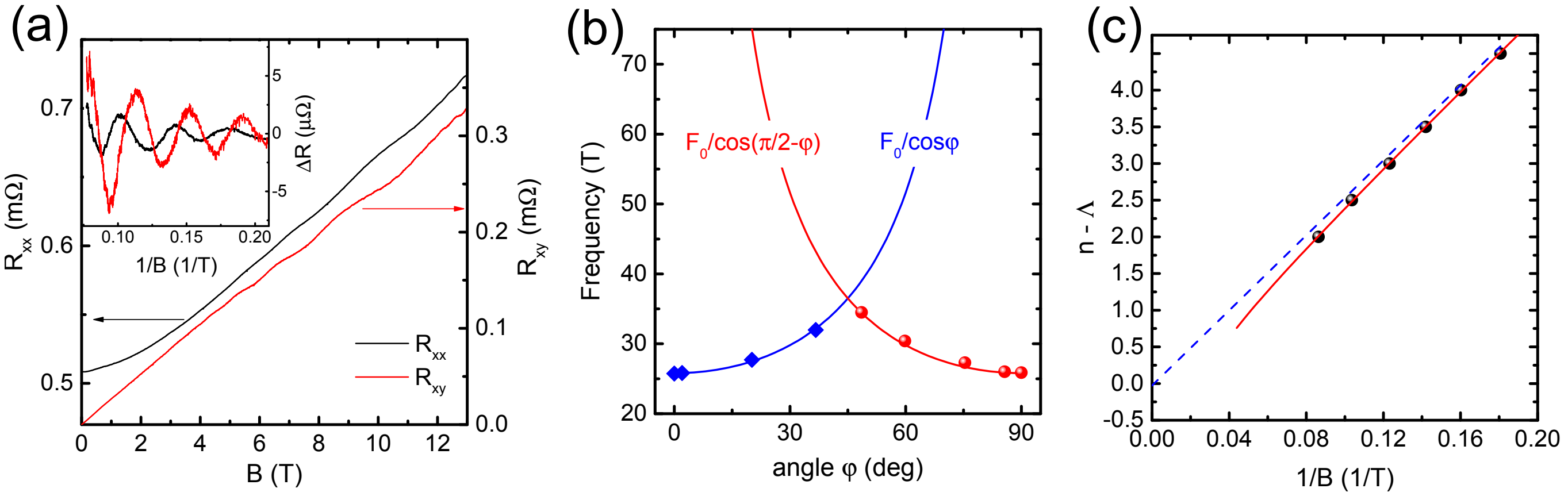

FIG. 1. The Shubnikov - de Haas and Quantum Hall oscillations of topological surface states in SnTe. (a) Raw $R_{xx}$ and $R_{xy}$ resistances versus magnetic field taken at $\boldsymbol{B} \parallel [001]$. Wavy lines are sums of bulk and surface contributions. Very small values of $R_{xy}$ are caused by the high hole density of bulk SnTe. Inset: Oscillatory components $\Delta R_{xx}$ and $\Delta R_{xy}$ obtained by subtracting smooth backgrounds. (c) Frequencies of SdH oscillations related to the topological surface states, as extracted from chirp Z-transform for different tilt angles. Blue diamonds - experiment. Blue line represents $F_0/\cos(\varphi)$ dependence related to the top and bottom surfaces. Red circles - experiment. Red line represents $F_0/\cos(\pi/2\text{-}\varphi)$ dependence for the side surfaces. (c) Landau-level index plot of $R_{xx}$ extrema related to the surface states versus 1/B. Nonlinear fit (see Ref. 21) yields the Berry phase of $\pi$ characteristic of the topological surface states.

To demonstrate that the topological surface states exist on *all* 2D boundaries of the SnTe sample we analyse the $R_{xx}$ resistance measured for a magnetic field tilted with respect to the [001] crystal direction. In this configuration the magnetic field has components normal to both neighboring surfaces and thus both of them contribute to magneto-resistance tensor. The oscillatory data were analyzed using the Chirp Z-transform [20]. This generalization of the discrete Fourier transform is often applied to data obscured by considerable noise. The results of our analysis, shown in Fig. 1b, are characterized by following important features. First, the frequency originating from one surface follows the $F_0/cos\varphi$ dependence. Second, one observes a contribution of the other surface following the $F_0/cos[(\pi/2)\text{-}\varphi]$ dependence. Third, the complete symmetry of angular dependences for the two surfaces shows that the top and side surfaces are characterized by identical states. The chirp analysis did not reveal any additional frequency at $\varphi = 0°$ or $\varphi = 90°$, which means that the opposite surfaces are equivalent.

We also determine the Berry phase of our TSS. In Fig. 1c we show the Landau-level index plot of SdH extrema versus $1/B$. According to the theory [21–23], such plots are nonlinear functions of $1/B$ at high fields (or low numbers $n$). The Berry phase is determined by the asymptotic limit of such a plot for low fields. Figure 1c indicates the best fit to the data (red solid line) as well as its asymptote (dashed blue line). The determined phase offset $\gamma$ is -0.03. Thus, within the experimental accuracy, we obtain $\gamma \approx 0$ which corresponds to the nontrivial Berry phase of π characteristic of the Dirac fermions [22, 23].

We independently investigate the de Haas-van Alphen oscillations of magnetization. The magnetization depends only on the density of states and is a simpler phenomenon than magneto-transport, the latter being also influenced by carriers' scattering. As in the case of transport, the magnetization is composed of bulk SnTe and TSS contributions. Our measurements were carried out on freshly cleaved rectangular SnTe samples, similar to those used in the electric transport studies. The raw magnetization data contain a smooth nonoscillatory background (see Fig. S7 [12]). After this background is subtracted, the oscillatory component has the form shown in Fig. 2a for three temperatures.

To understand the observed behavior one must account for the presence of SnTe bulk carrier reservoir in contact with the surface states. To describe quantitatively our magnetization data we need to know an explicit form for the Landau level energies in TCI. This form is well approximated for the valence band by the

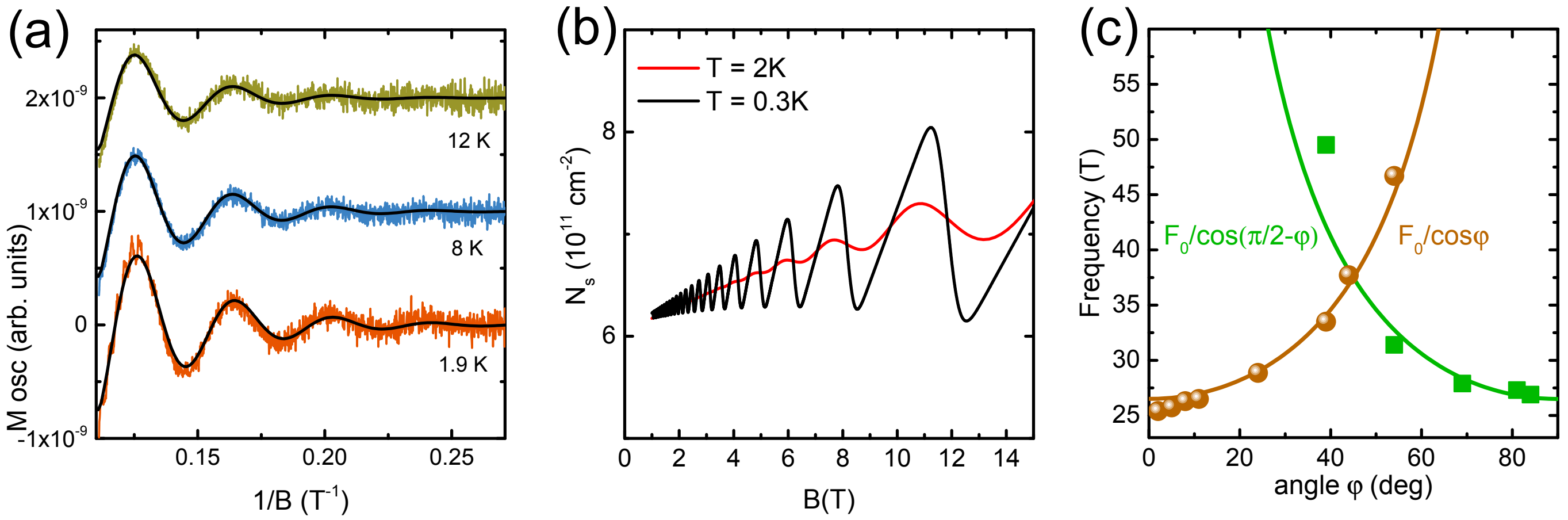

FIG. 2. Charge transfer and oscillatory magnetization of topological surface states in SnTe. (a) Oscillatory magnetization measured for the SnTe versus 1/B (the traces are shifted for clarity). Black solid lines are calculated for the Fermi energy pinned by the bulk reservoir using: $E_F$=80meV, $v_F$=4.4×10$^5$m/s, $\Gamma_0$=4.5meV, g*=57. (b) Oscillations of hole density in surface states calculated assuming that the bulk SnTe reservoir pins the Fermi energy of the system. (c) Frequencies of dHvA oscillations related to the TSS, as extracted from chirp Z-transform for different tilt angles of magnetic field. Brown circles -experiment. Brown line represents $F_0/\cos(\varphi)$ dependence related to the top and bottom surfaces. Green squares - experiment . Green line represents $F_0/\cos(\pi/2-\varphi)$ dependence for the side surfaces. For tilt angles 39 and 54 simultaneous contributions of two neighbouring surfaces are observed.

following formula, see [21]

$$E_n = \sqrt{2ne\hbar B v_F^2 + (g^*\mu_B B/2)^2}, \qquad (1)$$

where $n$ is the Landau level number, $e$ is the elementary charge, $h$ is the Planck constant, $\mu_B$ is the Bohr magneton and $g^*$ is the effective spin $g$-factor in the standard notation. The density of states (DOS) for a 2D system in the presence of a magnetic field is

$$\rho(E) = \frac{1}{2\pi L^2}\sum_n \sqrt{\frac{2}{\pi}}\frac{1}{\Gamma}\exp\left[-2\left(\frac{E-E_n}{\Gamma}\right)^2\right], \qquad (2)$$

where $E_n$ is given above, $L^2 = \hbar/eB$ and $\Gamma$ is the broadening of Gaussian peaks. The 2D electron density is given by 11

$$N_s = \int_0^\infty \frac{\rho(E)}{1+\exp\left[(E-E_F)/k_BT\right]}\,dE, \qquad (3)$$

where $E_F$ is the Fermi energy. As the magnetic field changes, the holes from the bulk SnTe reservoir go back and forth to the topological 2D surface states as illustrated in Fig. 2b. It is seen that changes of the hole density in TSS at T = 0.3 K are around 30%. It is sizable amount for TSS but negligible one for the reservoir of bulk SnTe. In principle, the Fermi energy in the bulk varies somewhat with the magnetic field but its changes are very slight [24]. In TSS we deal with a system characterized by a varying number of particles described by the grand canonical ensemble. Its potential $\Omega$ is

$$\Omega = -k_BT\int_{-\infty}^{+\infty}\rho(E)\log\left[1+\exp\left(\frac{E-E_F}{k_BT}\right)\right]dE. \qquad (4)$$

For semiconductor quantum wells in the presence of a quantizing magnetic field one usually considers two cases [13]. In the first, the number of carriers is constant and the Fermi energy oscillates. In the second, the Fermi energy is fixed by a reservoir and the number of carriers oscillates. In this case one has to take into account the fact that the electrostatic confining potential depends on the electric charge, so the system "breathes" when the carriers move back and forth [13]. Here, we deal with the third case in which the Fermi energy is fixed by the reservoir, the carriers also move back and forth, but since we do not deal with the confining potential, the system does not "breath". In our description we compute the grand potential $\Omega$ with the fixed value of the Fermi energy and then calculate the magnetization directly from the relation: $M = -\partial\Omega/\partial B|_{E_F}$.

The calculations were carried for the parameters indicated in the caption of Fig. 2a. From the frequency of dHvA oscillations we determine the hole density $N_s$ = 6.1×10$^{11}$ cm$^{-2}$ , and from the temperature dependence of their amplitudes the cyclotron effective mass $m^*$=0.08$m_0$ [12]. Using the relation between the Fermi wave-vector $k_F$and the density for spin-polarized 2D states: $k_F = \sqrt{4\pi N_s} = 2.7\times10^8\mathrm{m}^{-1}$, one obtains the Fermi velocity $v_F = \hbar k_F/m^* = 4\times10^5$m/s, which determines the linear Dirac dispersion of TSS. The above value agrees quite well with that determined in ARPES studies of SnTe [5]. The remaining parameters $\Gamma$ and $g^*$ are adjusted to fit the dHvA oscillations. We take into account the field dependence of the level broadening $\Gamma = \Gamma_0\sqrt{B}$ [25, 26]. As to the large value of $g^*$=57, it re-sults from the large spin-orbit interaction already manifested in the large bulk $g$-factors in SnTe [27, 28]. Figure 2a shows our oscillatory dHvA experimental results for magnetic field $B||[001]$ crystal direction and their theoretical description in the *1/B* scale. The frequency of dHvA oscillations is around 25.4 T, corresponding to the hole density of $N_s = 6.1{\times}10^{11}$ cm$^{-2}$. This is in agreement with the result of SdH oscillations obtained on a different sample ($N_s = 6.2{\times}10^{11}$ cm$^{-2}$). All in all, we reach a very good description of the oscillating magnetization of TSS at various temperatures thus confirming the charge transfer illustrated in Fig. 2b.

Next, similarly to the procedure applied to the transport effects we tilt the magnetic field in order to investigate the magnetization of the top and side surfaces together with their parallel counterparts. Figure S9a shows Chirp Z-transform power spectrum for different tilt angles [12]. Final results of our analysis are summarized in Fig. 2c which shows the dHvA frequencies obtained for the magnetic field tilted at different angles. These frequencies obey the same angle dependences as those followed by the SdH frequencies shown in Fig. 1b, which confirms that we deal as before with the top and bottom surfaces on the one hand and two side surfaces on the other. It is seen that for two angles: 39°and 54°one observes two different frequencies related to neighboring surfaces. Thus, the magnetization measurements fully confirm and augment the results described above for the transport data. In addition, as analyzed in [12], one obtains an identical π-Berry phase characteristic of the Dirac fermions.

The magnetization data shown in Fig. 2a exhibit for low values of *1/B* also high-frequency dHvA oscillations. They are better seen in a plot versus $B$, see Fig. S8 [12]. Analyzing these oscillations for tilted magnetic fields we identify them as originating from the 3D SnTe reservoir and determine the corresponding hole density to be $N_B = 1.4{\times}10^{20}$ cm$^{-3}$. Thus the magnetization data fully confirm also for the bulk reservoir the information obtained from magneto-transport. Thus, our data unambiguously indicate that, in spite of very high hole density in bulk SnTe and overlapping energies of the bulk and surface states, there exist completely independent sets of Landau levels in both subsystems.

It should be noted that in early papers on SnTe [29, 30] the authors observed at hole densities above $2.2{\times}10^{20}$ cm $^{-3}$ magneto-oscillations at the frequency range of 30-60 T, similar to our results shown in Fig. 1b and Fig. 2c. Since at that time the TSS were unknown, the observations were attributed to a second valence band at the Σ-points of the bulk Brillouin zone 25. However, we unambiguously demonstrate above with the use of the quantum Hall effect, angular dependence of SdH and dHvA oscillations and their $\pi$-Berry phase that we observe the 2D topological surface states revealed in the ARPES studies [5]. An additional indication that the results of [29, 30] should be attributed to the 2D states is that the bulk Σ- band in IV-VI rock-salt materials should be manifested by *two* sets of magneto-oscillations with strongly different frequencies [31], whereas the early data and our results show only single frequency.

In summary, it is experimentally demonstrated that as-grown bulk SnTe is *surrounded on all sides* by the metallic 2D topological surface states. We study surface and bulk states of topological crystalline insulator SnTe by investigating quantum magneto-transport and magnetization. Bulk and surface components are separated using oscillatory character of the studied effects. By tilting the external magnetic field we establish contributions of the topological states originating from neighboring surfaces perpendicular to each other. Correlation of quantum oscillations of the Hall and longitudinal magneto-resistances is observed. The magnetization data are described theoretically taking into account that the bulk SnTe reservoir pins the Fermi energy of the whole system. Landau-level index plots are used to determine the Berry phase of π in magneto-transport and magnetization oscillations confirming that the identified surface states are topological Dirac fermions. In all observations, presence of the large bulk reservoir of holes is strongly felt. The presented findings have universal character and apply to other topological crystalline insulators as well as to topological insulators.

We gratefully acknowledge elucidating discussions with Professor Tomasz Dietl. The work has been supported in part by the Polish National Science Centre grants 2012/07/B/ST3/03607 and 2014/15/B/ST3/03833.

* Krzysztof.Dybko@ifpan.edu.pl

[1] L. Fu, C. Kane, and E. Mele, Phys. Rev. Lett. **98**, 106803 (2007).

[2] Y. Ando and L. Fu, Annu. Rev. Condens. Matter Phys. **6**, 361 (2015).

[3] C. Bruene, C. X. Liu, E. G. Novik, E. M. Hankiewicz, H. Buhmann, Y. L. Chen, X. L. Qi, Z. X. Shen, S. C. Zhang, and L. W. Molenkamp, Phys. Rev. Lett. **106**, 1 (2011).

[4] Y. Xu, I. Miotkowski, and Y. P. Chen, Nat. Commun. **7**, 1 (2016).

[5] Y. Tanaka, Z. Ren, T. Sato, K. Nakayama, S. Souma, T. Takahashi, K. Segawa, and Y. Ando, Nat. Phys. **8**, 800 (2012).

[6] Y. Okada, M. Serbyn, H. Lin, D. Walkup, W. Zhou, C. Dhital, M. Neupane, S. Xu, Y. J. Wang, R. Sankar, F. Chou, A. Bansil, M. Z. Hasan, S. D. Wilson, L. Fu, and V. Madhavan, Science **341**, 1496 (2013).

[7] P. Sessi, D. Di Sante, A. Szczerbakow, F. Glott, S. Wilfert, H. Schmidt, T. Bathon, P. Dziawa, M. Greiter, T. Neupert, G. Sangiovanni, T. Story, R. Thomale, and M. Bode, Science **354**, 1269 (2016).

[8] T. H. Hsieh, H. Lin, J. Liu, W. Duan, A. Bansil, and L. Fu, Nat. Commun. **3**, 982 (2012).

[9] P. Dziawa, B. J. Kowalski, K. Dybko, R. Buczko, A. Szczerbakow, M. Szot, E. Łusakowska, T. Balasubramanian, B. M. Wojek, M. H. Berntsen, O. Tjernberg, and T. Story, Nat. Mater. **11**, 1023 (2012).

[10] S.-Y. Xu, C. Liu, N. Alidoust, M. Neupane, D. Qian, I. Belopolski, J. D. Denlinger, Y. J. Wang, H. Lin, L. a. Wray, G. Landolt, B. Slomski, J. H. Dil, a. Marcinkova, E. Morosan, Q. Gibson, R. Sankar, F. C. Chou, R. J. Cava, a. Bansil, and M. Z. Hasan, Nat. Commun. **3**, 1192 (2012).

[11] A. Szczerbakow and K. Durose, Prog. Cryst. Growth Charact. Mater. **51**, 81 (2005).

[12] See Supplemental Material at http://link.aps.org/ supplemental/ for additional experimental and technical details including Refs [32-35].

[13] W. Zawadzki, A. Raymond, and M. Kubisa, Phys. Status Solidi **251**, 247 (2014).

[14] M. Von Ortenberg, O. Portugall, W. Dobrowolski, A. Mycielski, R. Gałazka, and F. Herlach, J. Phys. C Solid State Phys. **21**, 5393 (1988).

[15] J. G. Analytis, R. D. McDonald, S. C. Riggs, J.-H. Chu, G. S. Boebinger, and I. R. Fisher, Nat. Phys. **6**, 960 (2010).

[16] M. Sasaki, N. Miyajima, H. Negishi, M. Inoue, V. A. Kulbachinskii, K. Suga, Y. Narumi, and K. Kindo, Phys. B **298**, 510 (2001).

[17] N. Miyajima, M. Sasaki, H. Negishi, M. Inoue, V. A. Kulbachinskii, A. Y. Kaminskii, and K. Suga, J. Low Temp. Phys. **123**, 219 (2001).

[18] D. Shoenberg, *Magnetic oscillations in metals* (Cambridge University Press, 1984).

[19] S. H. Simon and B. I. Halperin, Phys. Rev. Lett. **73**, 3278 (1994).

[20] L. Rabiner, R. Schafer, and C. Rader, IEEE Trans. Audio Electroacoust. **17**, 86 (1969).

[21] Z. Wang, Z.-G. Fu, S.-X. Wang, and P. Zhang, Phys. Rev. B **82**, 1 (2010).

[22] G. P. Mikitik and Y. V. Sharlai, Phys. Rev. B **85**, 5 (2012).

[23] A. R. Wright and R. H. McKenzie, Phys. Rev. B **87**, 1 (2013).

[24] J. S. Blakemore, *Semiconductor Statistics* (Pergamon Press, 1962).

[25] T. Ando and Y. Uemura, J. Phys. Soc. Japan **36**, 959 (1974).

[26] J. P. Eisenstein, H. L. Stormer, V. Narayanamurti, A. Y. Cho, A. C. Gossard, and C. W. Tu, Phys. Rev. Lett. **55**, 875 (1985).

[27] R. L. Bernick and L. Kleinman, Solid State Commun. **8**, 569 (1970).

[28] O. Pankratov and B. Volkow, in *Physics Reviews*, edited by I. Khalatnikov (Harwood Academic Publishers GmbH, Switzerland, 1987) pp. 357–446.

[29] J. R. Burke, R. S. Allgaier, B. B. Houston, J. Babiskin, and P. G. Siebenmann, Phys. Rev. Lett. **14**, 360 (1965).

[30] H. T. Savage, B. Houston, and J. R. Burke, Phys. Rev. B **6**, 2292 (1972).

[31] P. Giraldo-Gallo, B. Sangiorgio, P. Walmsley, H. J. Silverstein, M. Fechner, S. C. Riggs, T. H. Geballe, N. A. Spaldin, and I. R. Fisher, Phys. Rev. B **94**, 1 (2016).

[32] G. Springholz and G. Bauer, "Semiconductors, IV-VI," (2014).

[33] J. Xiong, Y. Luo, Y. Khoo, S. Jia, R. J. Cava, and N. P. Ong, Phys. Rev. B **86**, 1 (2012).

[34] T. Ando, A. B. Fowler, and F. Stern, Rev. Mod. Phys. **54**, 437 (1982).

[35] G. Nachtwei, D. Schulze, G. Gobsch, G. Paasch, W. Kraak, H. Krüger, and R. Herrmann, Phys. status solidi **148**, 349 (1988).

# Supplemental Material for

## Experimental evidence for topological surface states wrapping around bulk SnTe crystal

K. Dybko, M. Szot, A. Szczerbakow, M. U. Gutowska, T. Zajarniuk, J. Z. Domagala,

A. Szewczyk, T. Story and W. Zawadzki

*Institute of Physics, Polish Academy of Sciences,*
*Aleja Lotnikow 32/46*
*PL-02668 Warsaw, Poland*

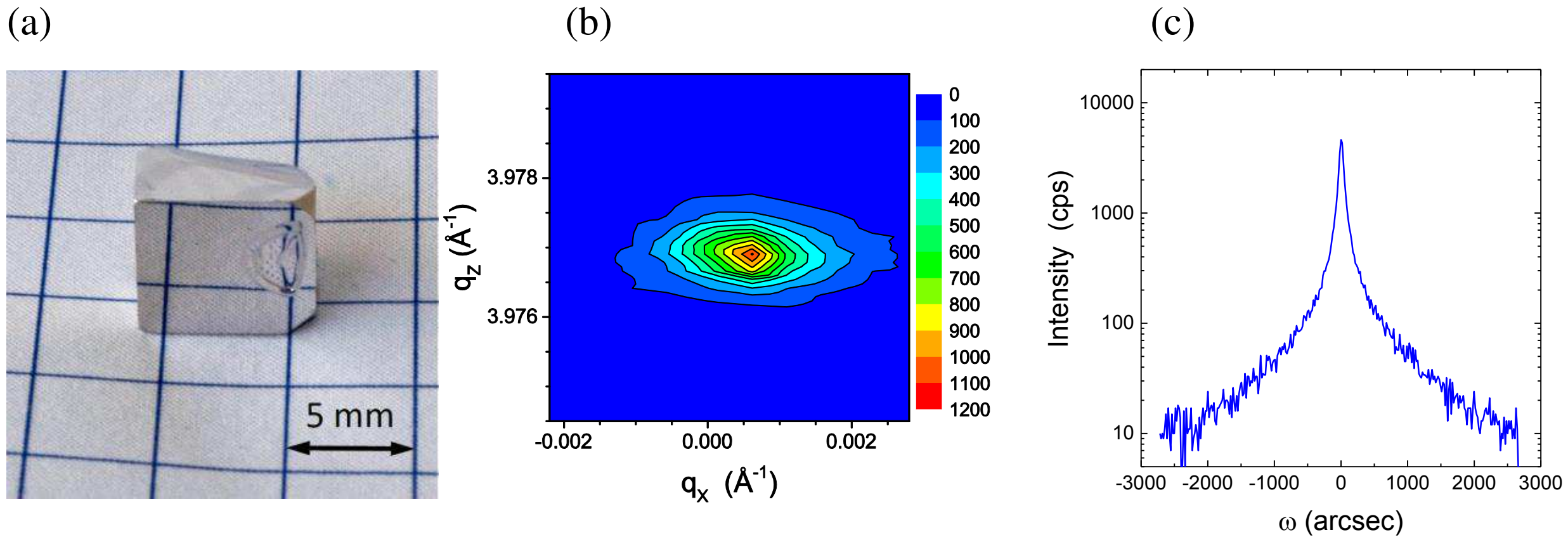

Fig. S1. (a) A single crystal of SnTe grown by self-selective vapour growth (SSVG) method. It is naturally terminated by equivalent (100) planes being the crystal cleavage planes. (b) Reciprocal space map of the 004 reflection shows distribution of the intensity around the Bragg nod 004 in counts per second units. (c) Rocking curve of the 004 reflection obtained in X-ray diffraction analysis. Full width at half maximum of the rocking curve peak equals to 80 arcsec confirming high crystallographic quality of the sample.

## 1. SnTe bulk material

The crystal growth of non-stoichiometric SnTe via the vapour phase required precise setting up the tin/tellurium ratio ensuring minimal total vapour pressure in the sealed growth ampoule, which is not achievable by weighing. Hence, post-synthesis annealing was applied. For an amount of approximately 80g Sn, an excessive portion of the more volatile Te was used to obtain the composition of $Sn_{0.49}Te_{0.51}$. Synthesis ampoule of 26mm bore and 250mm length was applied for melting the mixture at 950 °C. Next, the ampoule end containing the solidified material was placed in a furnace with 600 °C, while the empty end was sticking out to the room temperature for about 2 weeks to remove the elemental Te and also to set up the tin/tellurium proportion close to the metal-saturation status. Such status ensures minimal total vapour pressure optimal for the crystal growth. The growth was performed in the temperature profile ensuring radial material transport from neighbourhood of the ampoule walls to the surface of the source material. Condensation on the source material proceeded without contact with a foreign material and was marked by spontaneous selection of the crystals. After about a week crystal growth was completed [S1,S2]. The photograph of SnTe monocrystal is presented in Fig. S1 (a).

The structural quality of the samples was confirmed by X-ray diffraction method (different mode of scans, reciprocal space mapping) carried out with the use of Philips X'Pert MRD high resolution diffractometer (Cu tube, $\lambda$=1.54059 Å). The typical reciprocal space map of the 004 reflection (Fig. S1 (b)) confirms good crystal quality. Slight deformation of the node along $q_x$ axis was due to micro-mosaic defects structure. Lattice parameter of the investigated crystal is a=6. 3198±0.0001 Å.

The transport measurements were performed on samples freshly cleaved from the ingot. The typical dimensions were 5×1×0.4 $mm^3$. The electrical contacts were soldered with indium in a standard Hall bar geometry. The hole concentration determined by low field Hall effect at liquid helium temperature equals $3\times10^{20}$ $cm^{-3}$. The magneto-resistivity measurements were performed in magnetic fields up to 13 T and temperature range from 1.5 K to 20 K. The magnetic torque measurements were performed in Quantum Design PPMS with the base temperature 1.9 K. The magnetic field was set equidistant in 1/B between 3 T and 9 T producing 3000 total experimental points. The sample of dimensions 0.5×0.5×0.2 $mm^3$ was cleaved along (001) plane just before measurements to limit an influence of oxidation processes.

## 2. Magneto-resistance oscillations versus B

According to ref. [S3] in the regime of QHE the following rule is obeyed for the two resistivities: $R_{xx} = \alpha B \frac{dR_{xy}}{dB}$ , where $\alpha$ is a proportionality factor. However, according to the analysis presented in ref.[S4], when the surface conductance coexists with the large bulk conductance, the $R_{xx}$ maxima of the 2D resistance become minima. We take into account this exchange by reversing sign of $R_{xx}$ in Fig. S2. It can be seen that the above formula relating the two resistivities is quite well satisfied. This confirms again the 2D character of QHE contribution to the total Hall resistivity in Fig. 1 (a) of the main text.

We use the results for $R_{xx}$ in yet another context. It is known that, if $R_{xy}$>> $R_{xx}$, the resistance $R_{xx}$ is proportional to the conductance $G_{xx}$. On the other hand, the Shubnikov-de Haas oscillations of conductance in 2D systems are roughly proportional to the density-of-states (DOS) in a magnetic field at the Fermi energy [S5]. In consequence, we show in Fig. S3 again the experimental $R_{xx}$ data (black curve) together with the density of states (given by Eq. (2) in main text) at the Fermi energy. It is confirmed that the presence of highly conductive bulk reservoir causes the maxima and minima of 2D SdH oscillations to exchange their role. This coincides with the natural result shown in Fig. S3 that the maxima of SdH correspond to the maxima of DOS (note the reversed scale of $R_{xx}$ values). The above features plus the determined hole density of $6.2\times10^{11}$ $cm^{-2}$ unambiguously confirm that we deal with the surface 2D states.

The 3D regime of magneto-oscillations is seen at high values of B > 11T. In Fig. S4 we show $R_{xx}$ and $R_{xy}$ oscillations in this range of fields after subtracting the corresponding backgrounds. It is seen that this time the high frequency oscillations (F=322T) have the phase difference of $\pi$. According to the analysis of Nachtwei et al [S6] this corresponds to the 3D behaviour of magnetoresistances in which the oscillations of $R_{xy}$ are due to the contribution of oscillating conductance $G_{xx}$. This feature plus the determined hole density of $1.3\times10^{20}$ $cm^{-3}$ confirm that we observe in this range of fields the bulk 3D states.

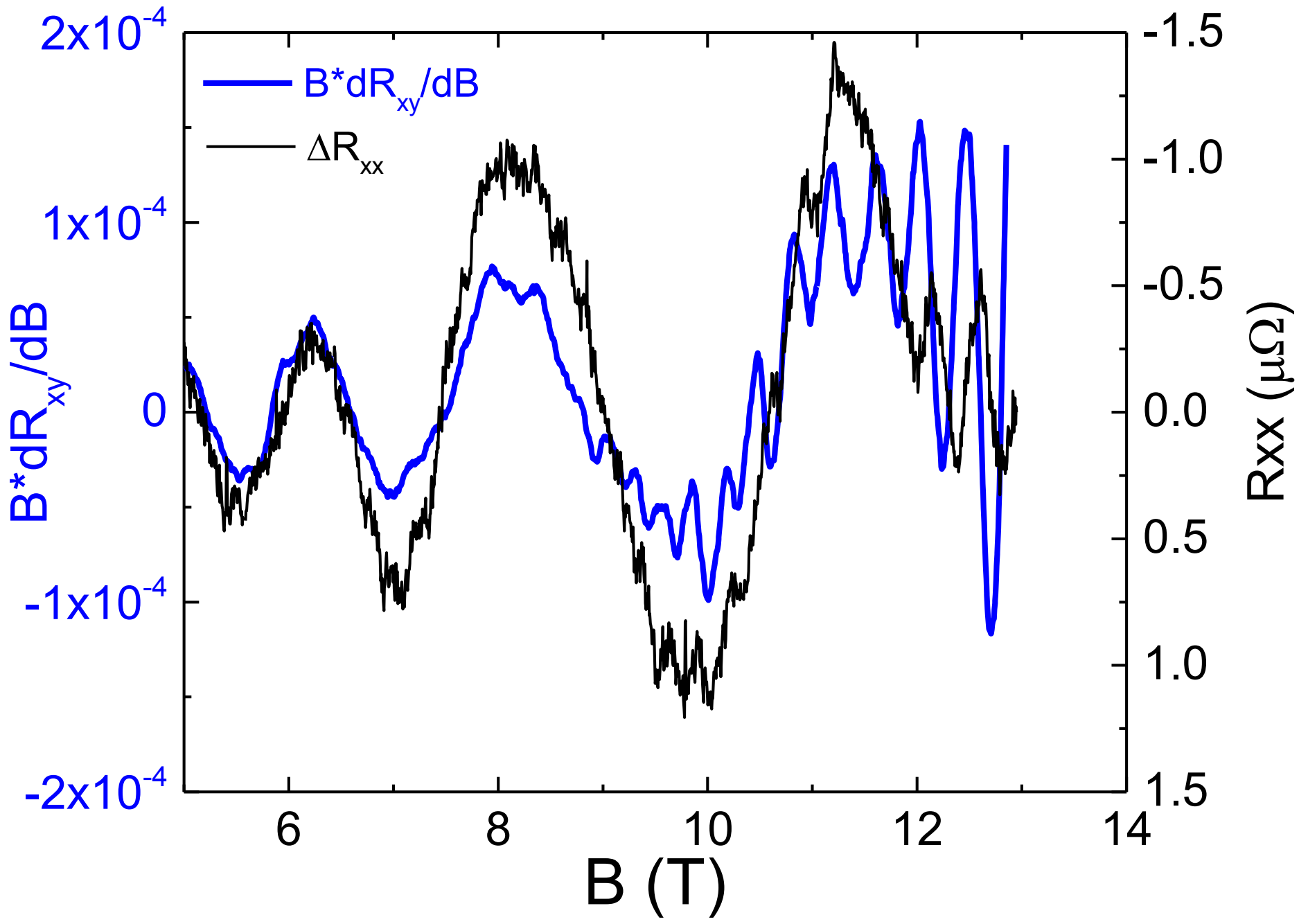

Fig. S2 Oscillatory components of experimental $R_{xx}$ and B•$dR_{xy}$/dB. Reversing the vertical scale of $R_{xx}$ caused by the presence of reservoir, see [S4], one obtains the agreement between the two quantities proposed in ref. [S3].

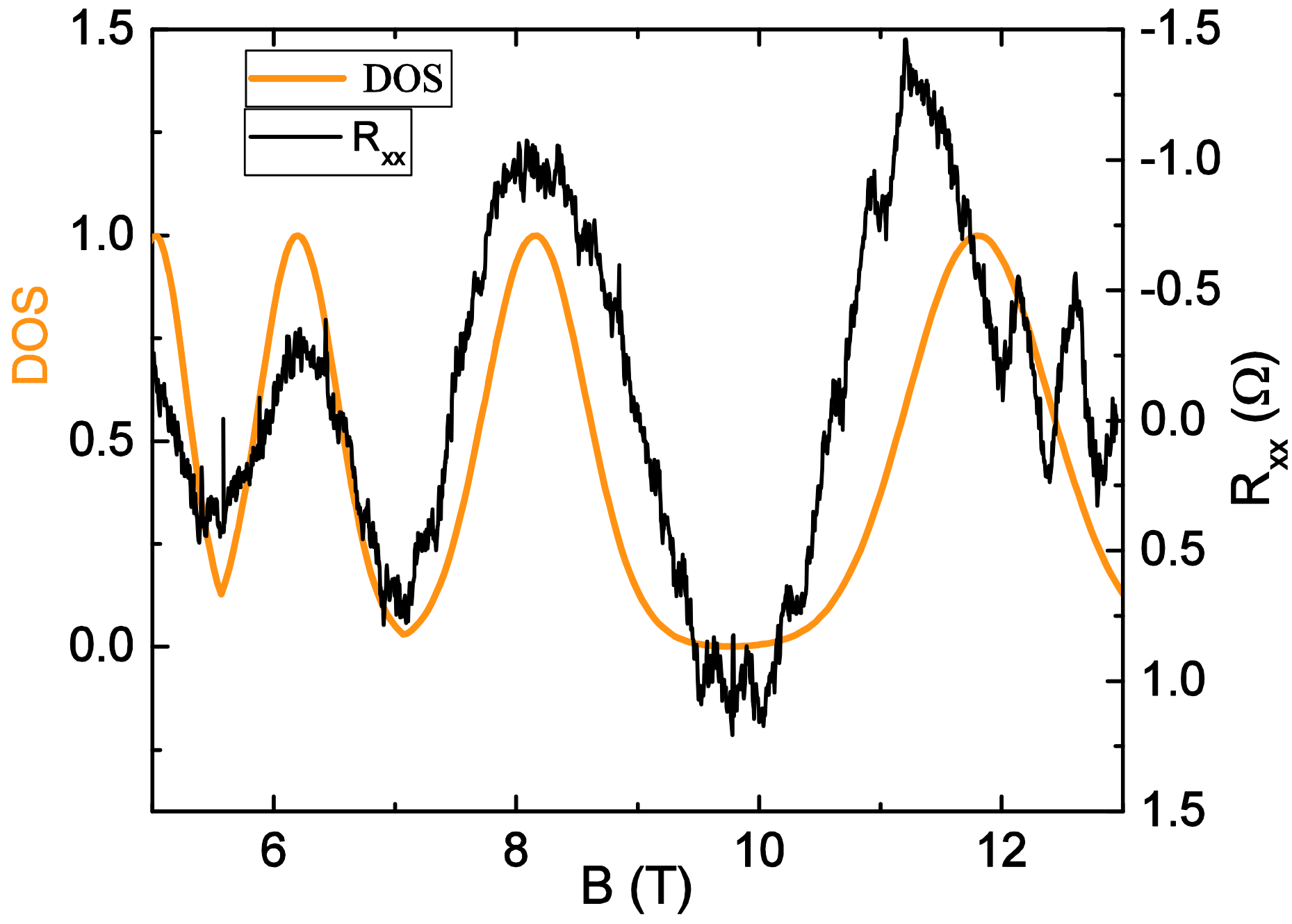

Fig. S3 Experimental Shubnikov-de Haas oscillations related to topological surface states (black trace) and calculated density of states at the Fermi energy pinned by the bulk reservoir. It is seen that the maxima of SdH after taking into account the change of phase due to the presence of reservoir (see ref. [S4]) correctly coincide with the maxima of DOS.

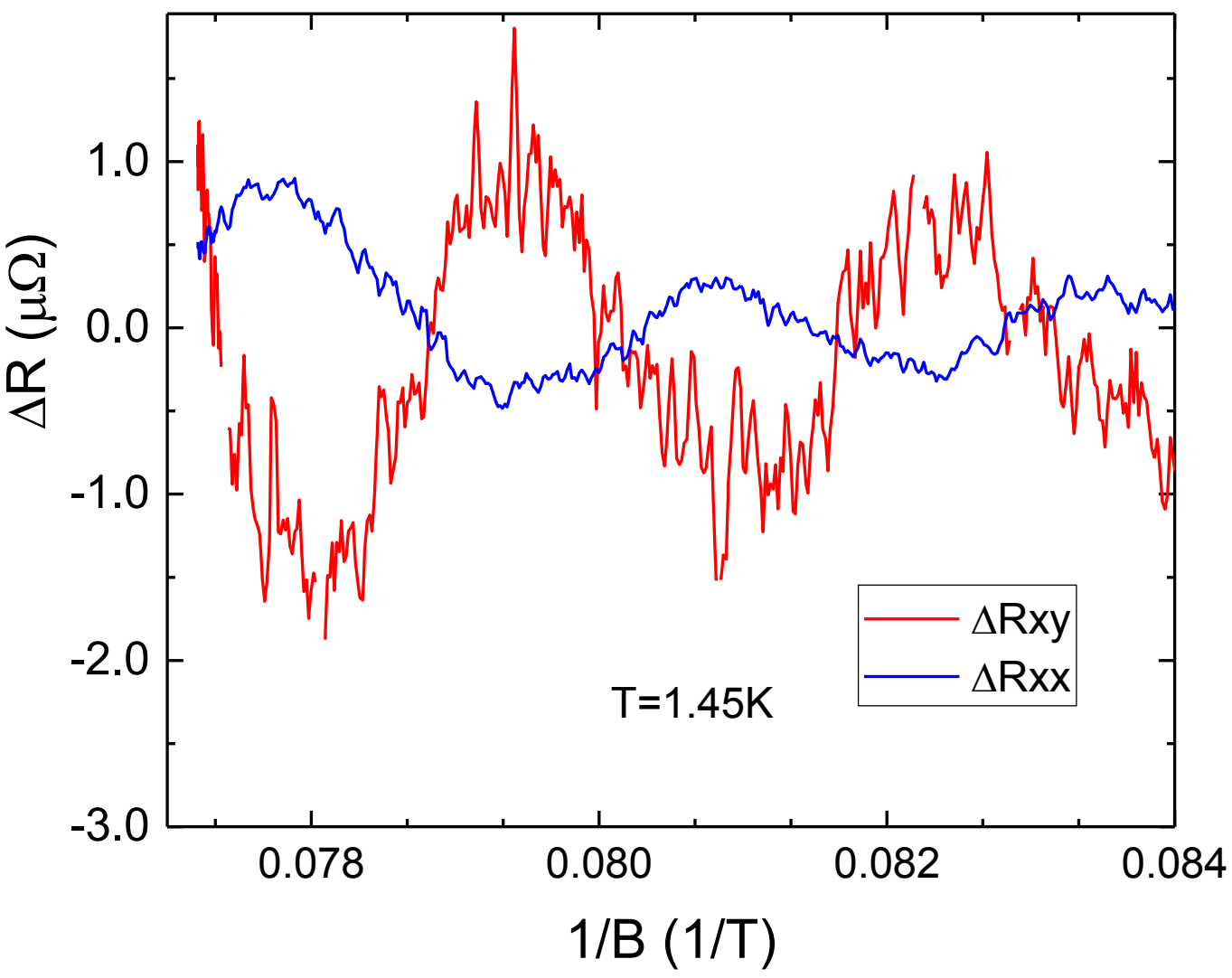

Fig. S4 Experimental oscillatory components of $R_{xx}$ and $R_{xy}$ magneto-resistances at higher fields 11T< B< 13T in 1/B scale. The phase difference of $\pi$ between two sets of oscillations indicates that one deals with 3D case in which oscillations of $R_{xy}$ are caused by the contribution of $G_{xx}$.

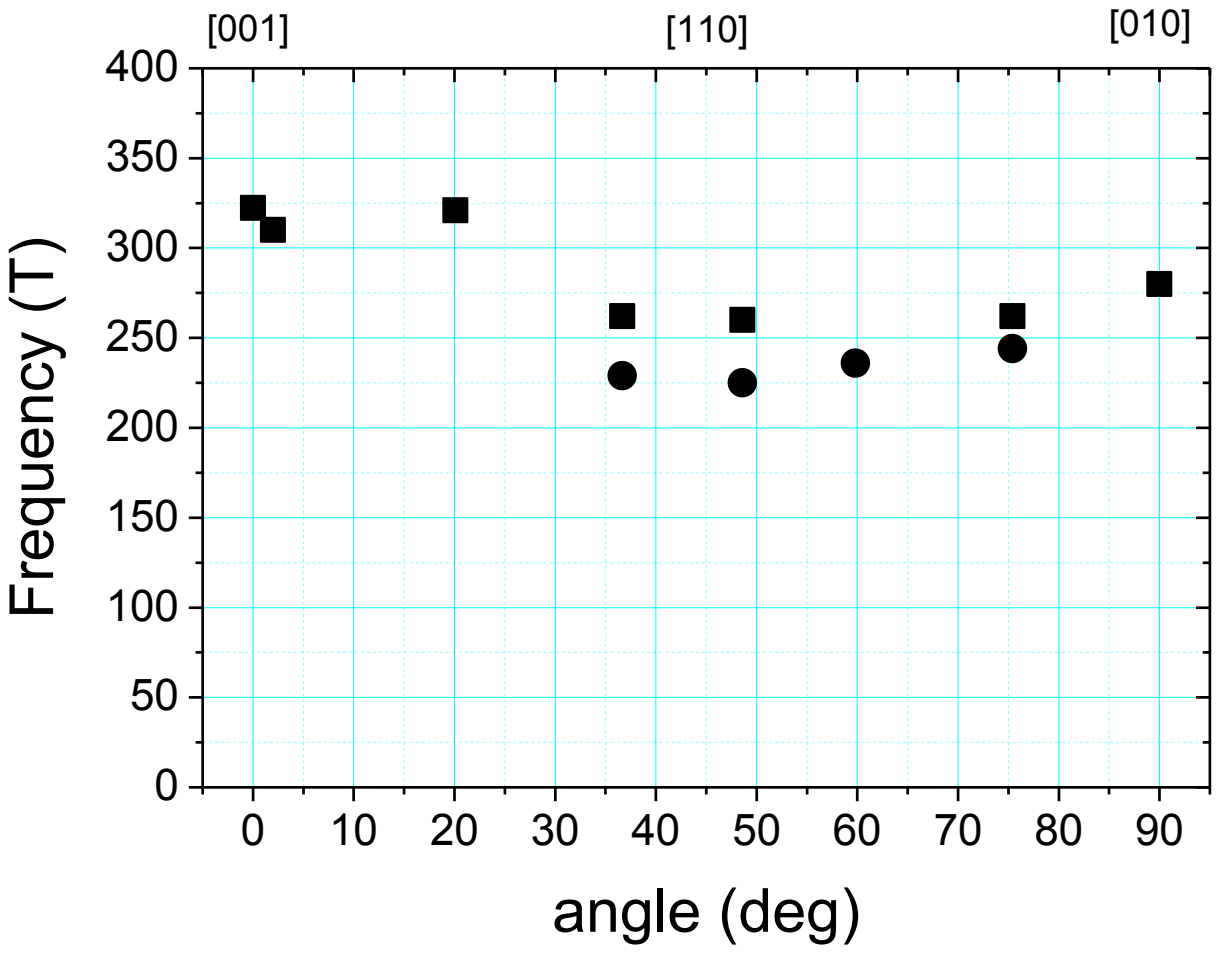

Figure S5 Angular dependence of high frequency SdH oscillations related to bulk reservoir.

The observed in high magnetic field region the high frequency oscillations of the bulk holes exhibit the angular dependence shown in Fig. S5. It is in good agreement with that reported in ref. [S7].

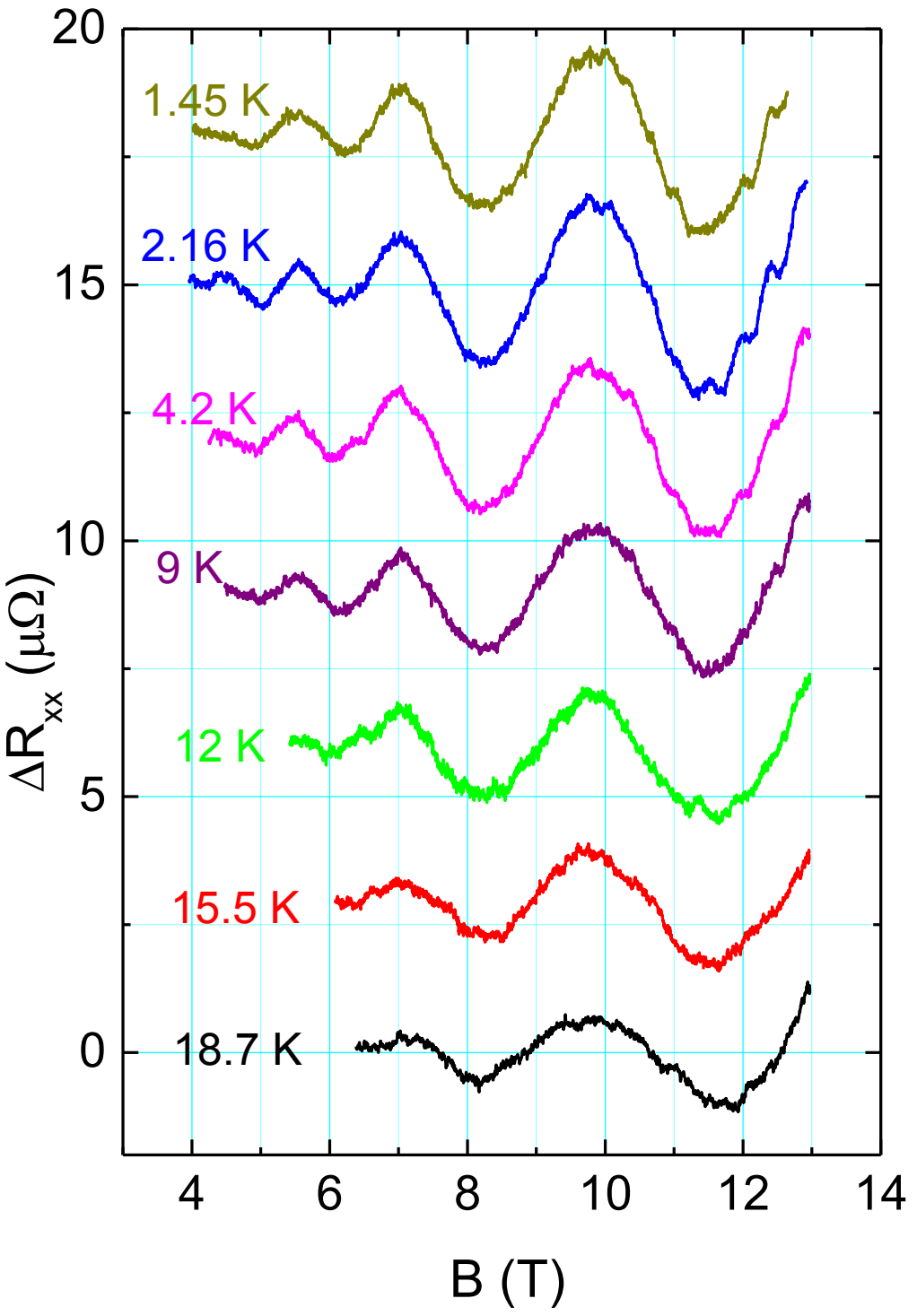

Fig. S6 Temperature dependence of SdH oscillations with smooth background removed. The traces are shifted by 3 $\mu\Omega$ for clarity.

## 3. Magnetization oscillations versus B

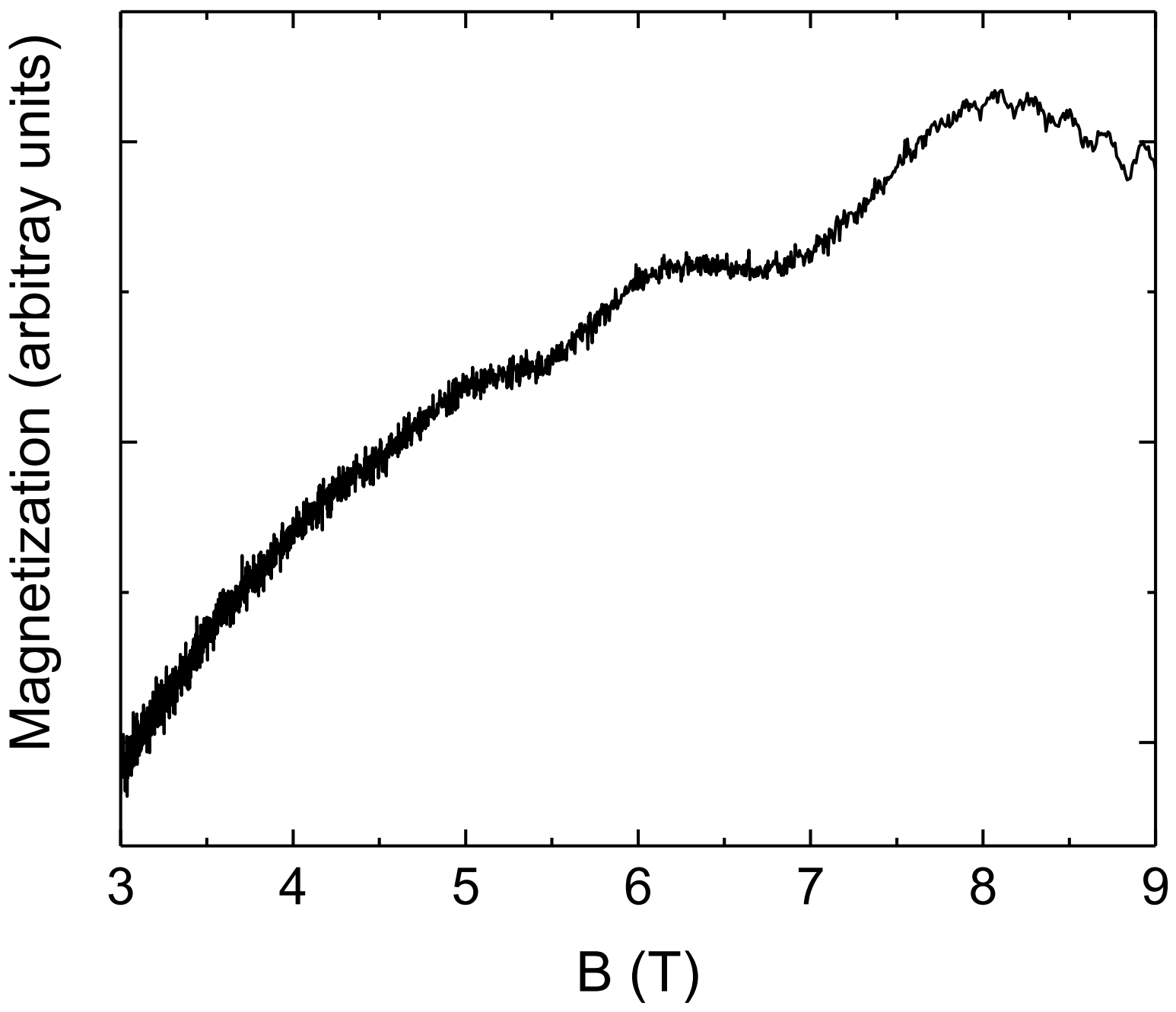

Fig. S7 The raw magnetization data versus magnetic field measured at lowest temperature T=1.9K.

Similarly to the plots of magneto-transport data, the magnetization oscillations related to bulk SnTe are better observed in the B scale. Below we show the high field region of dHvA oscillations in the B scale measured at four temperatures (Fig. S8). The oscillations, well seen at low temperatures, cease to be observable at T=8 K.

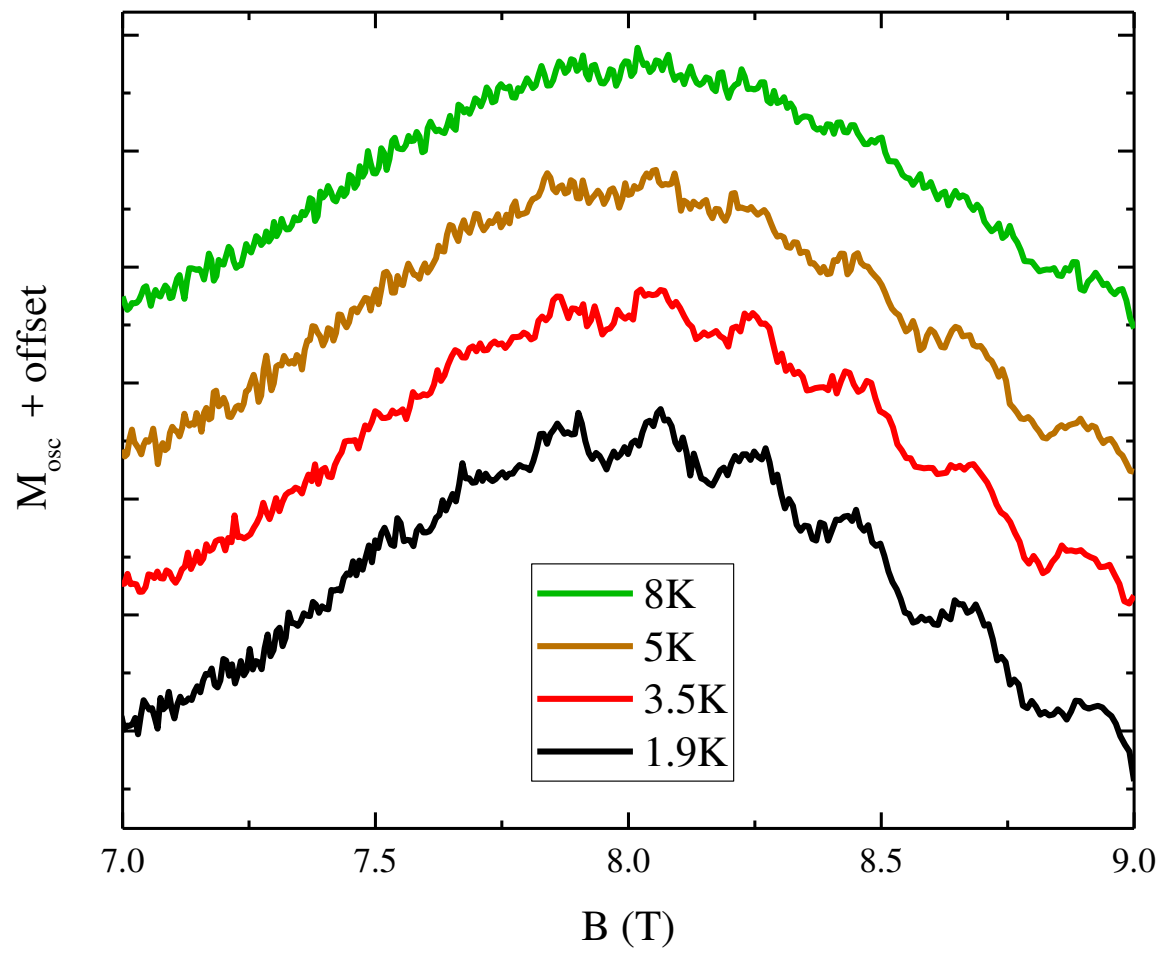

Fig. S8 The de Haas-van Alphen oscillations of magnetization related to the bulk reservoir, as observed at low temperatures in the high-field region.

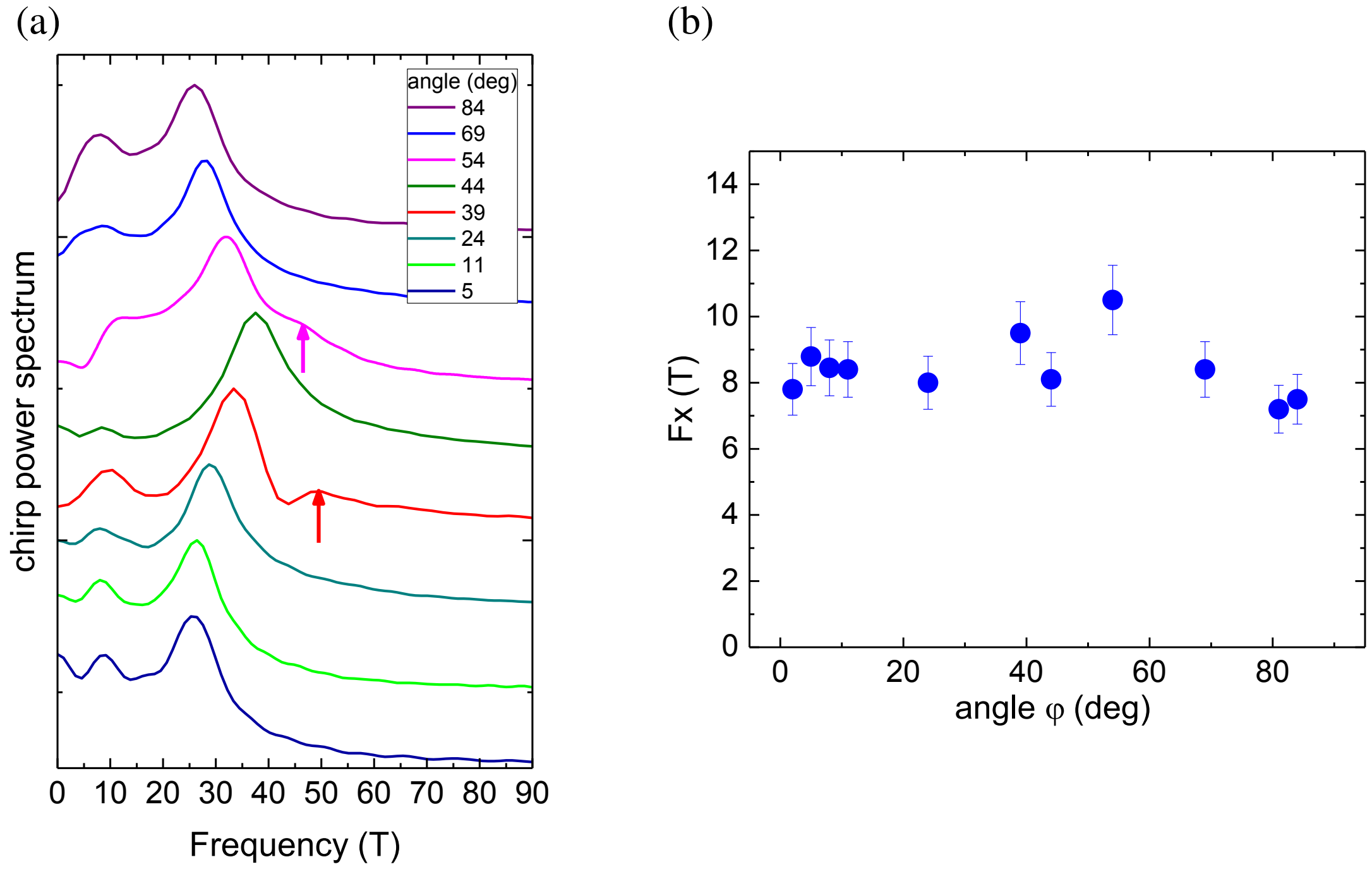

Fig. S9 (a) Power spectra obtained by Chirp Z- transformation of the dHvA oscillations for chosen tilt angles of magnetic field. For 39°and 54° field angles one observes simultaneously two peaks related to (001) and $(0\bar{1}0)$ surfaces. (b) The lowest frequency peak (around 8T) in the power spectrum of the de Haas-van Alphen oscillations shown in panel (a). The origin of the peak is at present not understood since its position does not exhibit any clear angular dependence. This means that it cannot be attributed to the surface states or to the 3D ellipsoidal bands of SnTe.

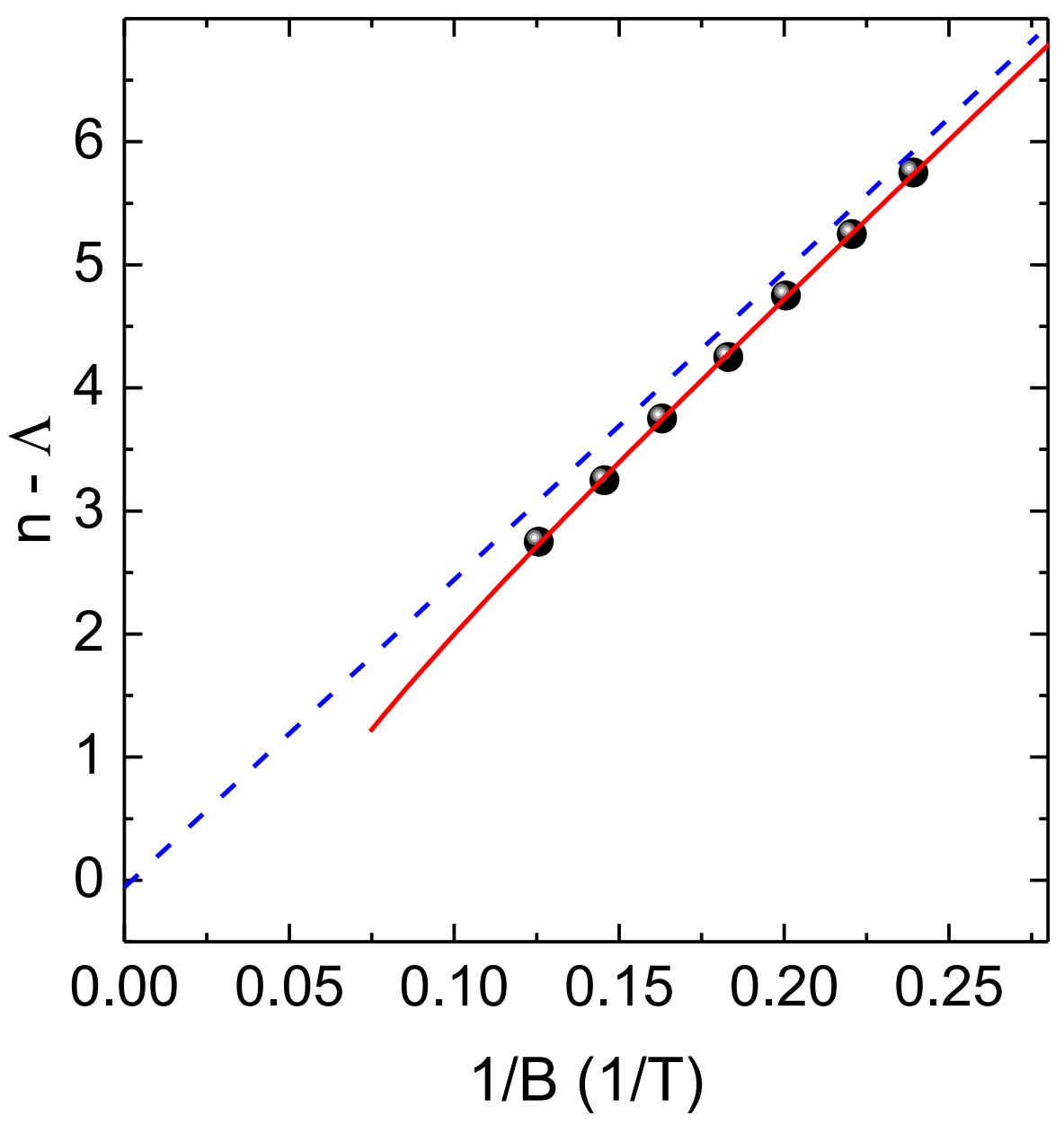

Fig. S10 The Landau-level index plot of oscillatory magnetization extrema related to the surface states versus 1/B. Λ=1/4 (3/4) attributed to maxima (minima). The red solid line represents nonlinear fit using the same procedure as presented in SdH case. The intercept of the asymptote of the fit (blue dashed line) with the y-axis gives phase offset γ. The determined γ is about 0, which corresponds to the nontrivial Berry phase of π characteristic of topological surface states.

## 4. Evaluation of effective mass and quantum mobility from quantum oscillatory effects SdH and dHvA

The SdH quantum oscillations of two dimensional spin polarized system in first harmonic approximation of the Lifshitz-Kosevich theory are expressed as:

$$\Delta R_{xx} \propto A \cdot \cos\left(2\pi\left(\frac{F}{B}+\gamma\right)\right) ,$$

F is frequency of oscillations in inverse magnetic field (1/B) and $\gamma$ is a phase factor containing information about Berry phase $\Phi_B$: $\gamma = 1/2 - \Phi_B/2\pi$. Thus $\gamma=1/2$ corresponds to normal Schroedinger system, while $\gamma=0$ corresponds to nontrivial Berry phase $\Phi_B=\pi$ characteristic for massless Dirac fermions. Amplitude $A=A_TA_D$ contains two damping factors, temperature damping $A_T = X/\sinh(X)$ , with $X=\alpha Tm^*/B$ and Dingle damping $A_D=\exp(-\pi/\mu_q)$ , where: $\alpha=14.69$ T/K , m*- cyclotron effective mass, $\mu_q$ - quantum mobility.

The oscillatory magnetization ($M_{osc}$), it is dHvA effect has the same functional dependence, provided cosine function is replaced with sine function.

The effective masses of topological surface states obtained from analysis of SdH and dHvA effects are equal to each other within experimental accuracy $(0.08\pm0.01)m_0$ and $(0.075\pm0.005)m_0$. Further analysis yielded quantum mobilities of TSS which again coincide within 20 percent accuracy.

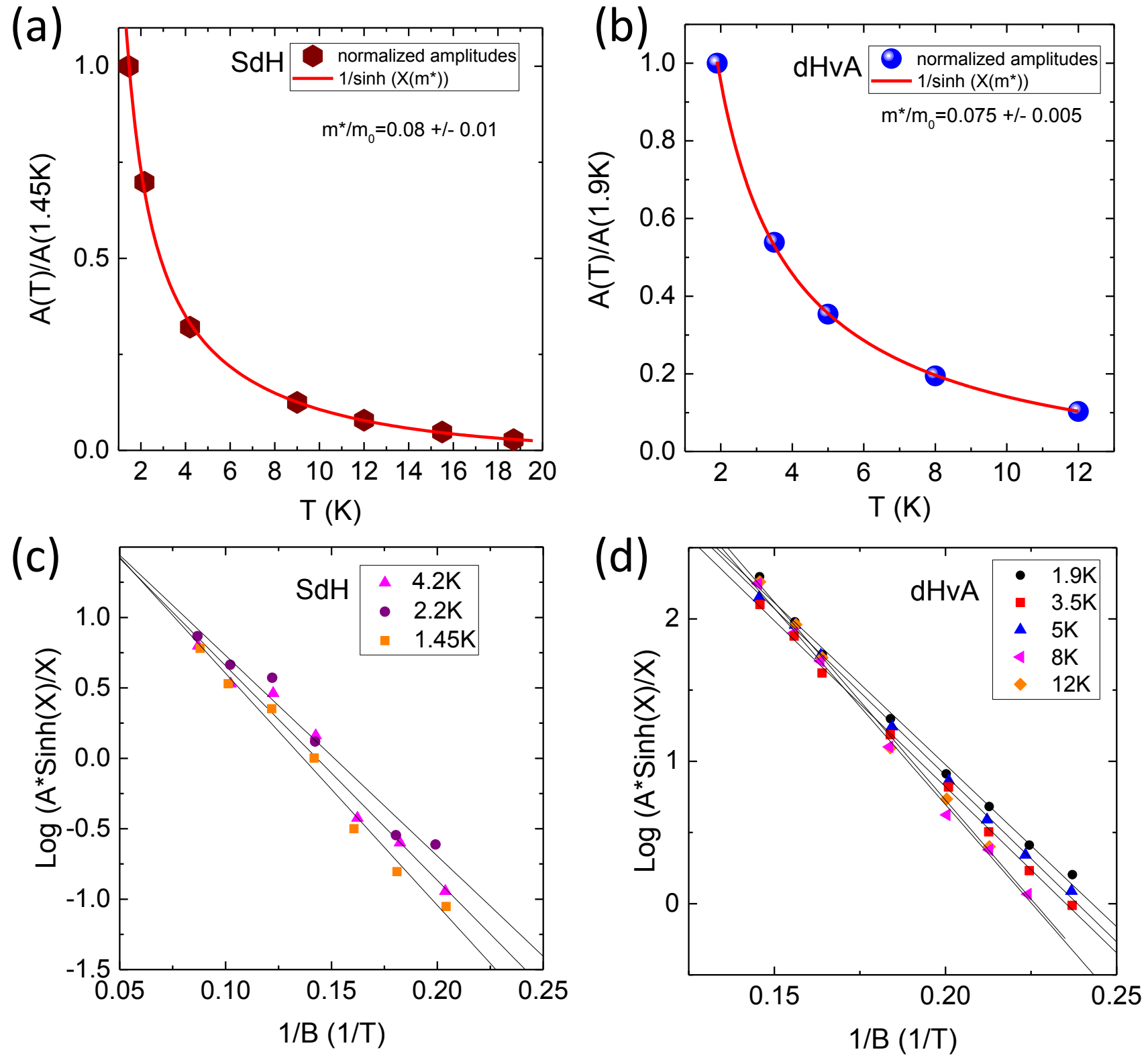

Fig. S11 (a) and (b) Normalized temperature-dependent amplitudes and Lifshitz-Kosevich fit for SdH and dHvA effect respectively. (c) and (d) Dingle plots for both SdH and dHvA effects obtained for different temperatures indicated in legend. The slope of linear fits gives $-\pi/\mu_q$, from which the quantum mobility can be calculated.

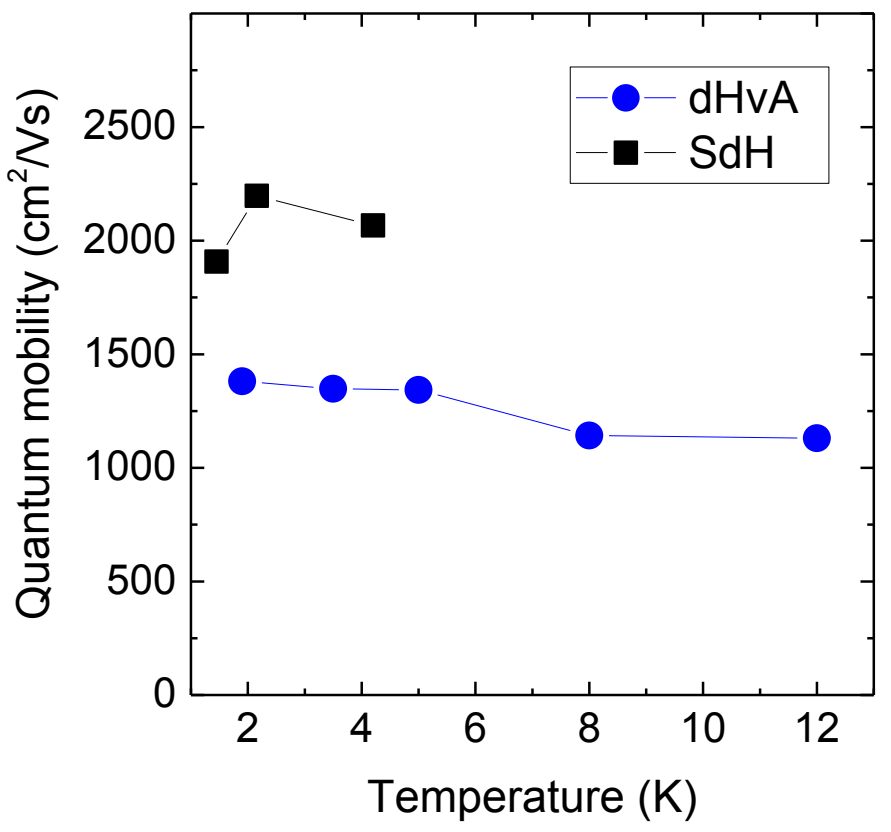

Fig. S12 Temperature dependence of quantum mobility of TSS obtained from Dingle analysis in Fig. S11 (c) and (d) for SdH and dHvA effects.

## References

[S1] G. Springholz and G. Bauer, Wiley Encycl. Electr. Electron. Eng. 1 (2014).

[S2] A. Szczerbakow and K. Durose, Prog. Cryst. Growth Charact. Mater. **51**, 81 (2005).

[S3] S. H. Simon and B. I. Halperin, Phys. Rev. Lett. **73**, 3278 (1994).

[S4] J. Xiong, Y. Luo, Y. Khoo, S. Jia, R. J. Cava, and N. P. Ong, Phys. Rev. B - Condens. Matter Mater. Phys. **86**, 1 (2012).

[S5] T. Ando, A. B. Fowler, and F. Stern, Rev. Mod. Phys. **54**, 437 (1982).

[S6] G. Nachtwei, D. Schulze, G. Gobsch, G. Paasch, W. Kraak, H. Krüger, and R. Herrmann, Phys. Status Solidi **148**, 349 (1988).

[S7] J. R. Burke, R. S. Allgaier, B. B. Houston, J. Babiskin, and P. G. Siebenmann, Phys. Rev. Lett. **14**, 360 (1965).